\documentclass[12pt,english]{article}
\usepackage[T1]{fontenc}
\usepackage[latin9]{inputenc}
\usepackage{amsmath}
\usepackage{amssymb}
\usepackage{graphicx}

\makeatletter
\usepackage{jheppub}
\def\@fpheader{\relax}
\allowdisplaybreaks[1]

\makeatother

\usepackage{babel}
\begin{document}
\subheader{}

\title{Deformed algebra and the effective dynamics of the interior of black holes}

\author[a]{Pasquale Bosso,}
\author[b]{Octavio Obreg\'on,}
\author[c]{Saeed Rastgoo,}
\author[b]{Wilfredo Yupanqui.}
\affiliation[a]{Department of Physics and Astronomy, University of Lethbridge,\protect\\ 4401 University Drive, Lethbridge, Alberta, Canada, T1K 3M4} 
\affiliation[b]{Departamento de F\'isica, Divisi\'on de Ciencias e Ingenier\'ias, Universidad de Guanajuato
Loma del Bosque 103, Le\'on 37150, Guanajuato, M\'exico.}
\affiliation[c]{Department of Physics and Astronomy, York University\\ 4700 Keele Street,Toronto, Ontario M3J 1P3, Canada.}
\emailAdd{pasquale.bosso@uleth.ca}
\emailAdd{octavio@fisica.ugto.mx}
\emailAdd{srastgoo@yorku.ca}
\emailAdd{w.yupanquicarpio@ugto.mx}

\abstract{We consider the classical Hamiltonian of the interior of the Schwarzschild black hole in Ashtekar-Barbero connection formalism.
Then, inspired by generalized uncertainty principle models, we deform the classical canonical algebra and derive the effective dynamics of the model under this modification.
We show that such a deformation leads to the resolution of the singularity of the black hole and a minimum nonzero radius for the infalling 2-spheres, provided that the deformation parameters are chosen to be negative. 
}
\maketitle

\section{Introduction}

Black holes, particularly their interior and singularity, are one
of the most important playgrounds of quantum gravity. Any full or
even effective theory of quantum gravity is expected to resolve black
hole singularities one way or the other. In Loop Quantum Gravity (LQG)
\cite{Thiemann:2007pyv} which is one of the main nonperturbative
approaches to the quantization of gravity, there has been numerous
studies of both the interior and the full spacetime of black holes
in four and lower dimensions \cite{Bojowald:2004af,Ashtekar:2005qt,Bojowald:2005cb,Bohmer:2007wi,Boehmer:2008fz,Corichi:2015xia,BenAchour:2017ivq,Ashtekar:2018cay,BenAchour:2018khr,Alesci:2018loi,Barrau:2018rts,Alesci:2019pbs,Aruga:2019dwq,BenAchour:2020gon,Bodendorfer:2019cyv,Bodendorfer:2019nvy,Bojowald:2008bt,Bojowald:2008ja,Bojowald:2016itl,Bojowald:2016vlj,Bojowald:2018xxu,Brahma:2014gca,Campiglia:2007pb,Chiou:2008nm,Corichi:2015vsa,Cortez:2017alh,Gambini:2008dy,Gambini:2009ie,Gambini:2011mw,Gambini:2013ooa,Gambini:2020nsf,Husain:2004yz,Husain:2006cx,Kelly:2020lec,Kelly:2020uwj,Kreienbuehl:2010vc,Modesto:2005zm,Modesto:2009ve,Olmedo:2017lvt,Thiemann:1992jj,Zhang:2020qxw,Ziprick:2016ogy,Campiglia:2007pr,Gambini:2009vp,Rastgoo:2013isa,Corichi:2016nkp,Morales-Tecotl:2018ugi,BenAchour:2020bdt,BenAchour:2020mgu,Gambini:2011nx}.
In such works, the classical Hamiltonian of the system is written
in terms of Ashtekar-Barbero connection and its conjugate momentum,
which is the densitized triad.
In the majority of the works focused on the interior of the black hole, such Hamiltonian is then quantized using a certain type of representation called polymer quantization \cite{Ashtekar:2002sn,Corichi:2007tf,Morales-Tecotl:2016ijb,Tecotl:2015cya,Flores-Gonzalez:2013zuk}.
This quantization procedure introduces a parameter into the theory, called the polymer scale, which sets a minimal scale of the model and determines the
onset of quantum gravitational effects.
These works show a general effective way of avoiding the singularity and furthermore introduce a bounce from a black hole to a white hole in the vacuum case.
Polymer quantization, in fact, introduces a certain modification the to algebra of the theory at the quantum level which can also be considered as an effective modification of the classical algebra.

There are other types of theories and models which consider different ways of
deforming the canonical algebra.
Among them is the
Generalized Uncertainty Principle (GUP).
Such a model has been introduced to phenomenologically account for a minimal measurable length in quantum mechanics \cite{Kempf:1994su,Ali:2011fa,Bosso:2017hoq,Bosso:2020aqm}.
In fact, many theories of quantum gravity and \emph{gedankenexperimente} in black hole physics predict a deviation from the Heisengerg uncertainty principles at high energies/momenta \cite{Gross:1987ar,Amati:1988tn,Maggiore:1993rv,Rovelli:1994ge,Garay:1994en,AmelinoCamelia:2008qg}.
On the other hand, GUP can be understood as a different quantization procedure which imposes a commutator between generalized position and momentum different from $i \hbar$ producing a minimal uncertainty in position or momentum. \cite{Bosso:2018ckz,Bosso:2019ljf}.
In a similar manner, models characterized by a minimal uncertainty in momentum, rather than in position, have been introduced.
Such models are termed Extended Uncertainty Principle (EUP) \cite{Bolen:2004sq,Park:2007az,Bambi:2007ty,Mignemi:2009ji,Mureika:2018gxl}.
More in general, one often refers to such modified models as Generalized Uncertainty Relations (GURs).
Here, we consider a modification of the Poisson brackets between generalized coordinates and momenta so to realize a minimal momentum.
Thus, our model is closer in spirit to EUP, although we will use tested techniques from GUP.

It is worth emphasizing that the strategy proposed in the present work is not unique.
In fact, as shown for example in \cite{Bishop:2019yft,Bishop:2020cep}, it is possible to consider representations for the position and momentum variables in the quantum sectors which keep the standard commutation relations.
However, as mentioned above, in the present work we considered the case in which only the momentum variables are modified, following the idea proposed in the seminal paper \cite{Kempf:1994su} and further elaborations \cite{Ali:2011fa,Bosso:2020aqm}.
Such approach is similar in nature to that considered for example in \cite{Mignemi:2011wh,Pramanik:2013zy,Pramanik:2014zfa,Chaichian:2014qba,Bosso:2018uus}.
A further alternative approach to the problem of a minimal uncertainty consists of smearing fields or wavefunctions so that they intrinsically exhibit an uncertainty in the quantities of interest regulated by geometrical considerations \cite{Lake:2018zeg,Lake:2020rwc}.
Finally, the same problem has been approached by other authors on a more heuristic level as well, with no requirement as for the representation of operators or commutators \cite{Scardigli:1999jh,Scardigli:2014qka,Blasone:2019wad,Casadio:2020rsj}.

In the present work, we consider the classical Hamiltonian of the interior of the Schwarzschild black hole in Ashtekar-Barbero connection formalism.
However, instead of applying the loop/polymer quantization techniques, we use a modification of the classical algebra inspired by GUP.
This way, we obtain similar qualitative effects in our model, including effective resolution of the singularity and a minimum nonzero radius for the infalling 2-spheres. 

This paper is organized as follows. In Sec. \ref{sec:Schw-int}, we
review the interior of the Schwarzschild black hole and its description
in terms of adapted Ashtekar-Barbero variables. In Sec. \ref{sec:Classical-dynamics},
a summary of the classical dynamics of the the interior in terms of
such variables is presented. Sec. \ref{subsec:GUP-redef-PB} presents
the general GUP-inspired algebra deformation that we use in this paper.
Using a certain type of such a deformation, we modify the classical
algebra in Sec. \ref{subsec:GUP-modified-dynamics} and derive the
resulting effective dynamics, and discuss its consequences including
the resolution of the classical singularity and a minimum nonzero radius for the infalling 2-spheres. In Sec. \ref{subsec:depend_L0} we address the dependence of physical quantities on the fiducial length of the homogeneous model and propose a prescription to solve this issue which leads to the introduction of a minimum physical length $\ell_c$ of the model. In Sec. \ref{subsec:beta-c-pc} we find a minimum nonzero radius for the infalling 2-spheres for our model. Furthermore, by identifying this radius with the one derived in LQG, we find an expression for $\ell_{c}$ in terms of the minimum area in LQG.
In Sec. \ref{subsec:Modif-Horiz}, we discuss the modifications to the
behavior at the horizon due to our effective model and make some brief comments on that.
Finally in Sec.
\ref{sec:Conclusion-and-outlook}, we present our concluding remarks
and discuss possible future routes.

\section{The interior of the Schwarzschild black hole\label{sec:Schw-int}}

It is well-known that upon crossing the event horizon of the Schwarzschild
black hole which is located at $R_{s}=2GM$ with $G$ being the Newton's
constant, the timelike and spacelike curves switch their causal nature.
Hence, given the metric of such a black hole 

\begin{equation}
ds^{2}=-\left(1-\frac{2GM}{r}\right)dt^{2}+\left(1-\frac{2GM}{r}\right)^{-1}dr^{2}+r^{2}\left(d\theta^{2}+\sin^{2}\theta d\phi^{2}\right),
\end{equation}
with $r\in(0,\infty)$ being the radial coordinate and the radius
of the 2-spheres in Schwarzschild coordinates $\left(t,r,\theta,\phi\right)$,
one can obtain the interior metric by switching $t\leftrightarrow r$
as 
\begin{equation}
ds^{2}=-\left(\frac{2GM}{t}-1\right)^{-1}dt^{2}+\left(\frac{2GM}{t}-1\right)dr^{2}+t^{2}\left(d\theta^{2}+\sin^{2}\theta d\phi^{2}\right).\label{eq:sch-inter}
\end{equation}
Here and throughout the paper, $t$ is the Schwarzschild time coordinate
(in the exterior) which has a range $t\in(0,2GM)$ in the interior.
Such a metric is a special case of a Kantowski-Sachs cosmological
spacetime that is given by the metric \cite{Collins:1977fg}
\begin{align}
ds_{KS}^{2}= & -N(T)^{2}dT^{2}+g_{xx}(T)dx^{2}+g_{\theta\theta}(T)d\theta^{2}+g_{\phi\phi}(T)d\phi^{2}\nonumber \\
= & -d\tau^{2}+g_{xx}(\tau)dx^{2}+g_{\Omega\Omega}(\tau)d\Omega^{2}.\label{eq:K-S-gener}
\end{align}
Note that $x$ here is not necessarily the radius $r$ of the 2-spheres
with area $A=4\pi r^{2}$. Also $\tau$ is the proper time.

The transformation between the two metrics (\ref{eq:sch-inter}) and
(\ref{eq:K-S-gener}) is given by 
\begin{equation}
d\tau^{2}=N(T)^{2}dT^{2}=\left(\frac{2GM}{t}-1\right)^{-1}dt^{2}.
\end{equation}
Note that $T$ is a generic time corresponds to the foliation with
lapse $N(T)$. Thus $T$ is in general different from the Schwarzschild
time $t$. The metric (\ref{eq:K-S-gener}) represents a spacetime
with spatial homogeneous but anisotropic foliations. A quick way to
see that is that $g_{xx}(\tau)$ and $g_{\Omega\Omega}(\tau)$ can
be considered as two distinct scale factors that affect the radial
and angular parts of the metric separately. As is evident from (\ref{eq:K-S-gener}),
such a system is a minisuperspace model due to incorporating a finite
number of configuration variables. Furthermore, it can be seen that
the spatial hypersurfaces have topology $\mathbb{R}\times\mathbb{S}^{2}$,
and the spatial symmetry group is the Kantowski-Sachs isometry group
$\mathbb{R}\times SO(3)$. Due to this topology with a noncompact
direction, $x\in\mathbb{R}$ in space, the symplectic form $\int_{\mathbb{R}\times\mathbb{S}^{2}}\text{d}^{3}x\,\text{d}q\wedge\text{d}p$
diverges. Therefore, one needs to choose a finite fiducial volume
over which this integral is calculated \cite{Ashtekar:2005qt}. This
is a common practice in the study of homogeneous minisuperspace models.
Here one introduces an auxiliary length $L_{0}$ to restrict the noncompact
direction to an interval $x\in\mathcal{I}=[0,L_{0}]$. The volume
of the fiducial cylindrical cell in this case is $V_{0}=a_{0}L_{0}$,
where $a_{0}$ is the area of the 2-sphere $\mathbb{S}^{2}$ in $\mathcal{I}\times\mathbb{S}^{2}$. 

In order to obtain the Hamiltonian of this system in connection variables,
one first considers the full Hamiltonian of gravity written in terms
of (the curvature) of the $su(2)$ Ashtekar-Barbero connection $A_{a}^{i}$,
and its conjugate momentum, the densitized triad $\tilde{E}_{a}^{i}$.
Using the Kantowski-Sachs symmetry, these variables can be written
as \cite{Ashtekar:2005qt}
\begin{align}
A_{a}^{i}\tau_{i}dx^{a}= & \frac{c}{L_{0}}\tau_{3}dx+b\tau_{2}d\theta-b\tau_{1}\sin\theta d\phi+\tau_{3}\cos\theta d\phi,\label{eq:A-AB}\\
\tilde{E}_{i}^{a}\tau_{i}\partial_{a}= & p_{c}\tau_{3}\sin\theta\partial_{x}+\frac{p_{b}}{L_{0}}\tau_{2}\sin\theta\partial_{\theta}-\frac{p_{b}}{L_{0}}\tau_{1}\partial_{\phi},\label{eq:E-AB}
\end{align}
where $b$, $c$, $p_{b}$ and $p_{c}$ are functions that only depend
on time, and $\tau_{i}=-i\sigma_{i}/2$ are a $su(2)$ basis satisfying
$\left[\tau_{i},\tau_{j}\right]=\epsilon_{ij}{}^{k}\tau_{k}$, with
$\sigma_{i}$ being the Pauli matrices. Substituting these into the
full Hamiltonian of gravity written in Ashtekar connection variables,
one obtains the symmetry reduced Hamiltonian constraint adapted to
this model as \cite{Ashtekar:2005qt}
\begin{equation}
H=-\frac{N\mathrm{sgn}(p_c)}{2G\gamma^{2}}\left[2bc\sqrt{|p_{c}|}+\left(b^{2}+\gamma^{2}\right)\frac{p_{b}}{\sqrt{|p_{c}|}}\right],\label{eq:H-class-N}
\end{equation}
while the diffeomorphism constraint vanishes identically due to homogenous
nature of the model. Here $\gamma$ is the Barbero-Immirzi parameter
\cite{Thiemann:2007pyv}.

Using symmetry of the model, its reduced symplectic form becomes \cite{Ashtekar:2005qt}
\begin{equation}
\boldsymbol{\Omega}=\frac{1}{2G\gamma}\left(dc\wedge dp_{c}+2db\wedge dp_{b}\right)
\end{equation}
from which one can read off the reduced Poisson brackets to be
\begin{equation}
\{c,p_{c}\}=2G\gamma,\quad\quad\{b,p_{b}\}=G\gamma.\label{eq:classic-PBs-bc}
\end{equation}
By substituting (\ref{eq:A-AB}) and (\ref{eq:E-AB}), and the components
of the inverse of the metric (\ref{eq:K-S-gener}), into the relation
between the inverse triad and the spatial metric $q_{ab}$,
\begin{equation}
qq^{ab}=\delta^{ij}\tilde{E}_{i}^{a}\tilde{E}_{j}^{b},
\end{equation}
one obtains for the generic metric (\ref{eq:K-S-gener}) adapted to
(\ref{eq:A-AB}) and (\ref{eq:E-AB})
\begin{align}
g_{xx}\left(T\right)= & \frac{p_{b}\left(T\right)^{2}}{L_{0}^{2}|p_{c}\left(T\right)|},\label{eq:grrT}\\
g_{\theta\theta}\left(T\right)= & \frac{g_{\phi\phi}\left(T\right)}{\sin^{2}\left(\theta\right)}=g_{\Omega\Omega}\left(T\right)=|p_{c}\left(T\right)|.\label{eq:gththT}
\end{align}
Note that the lapse $N(T)$ is not determined and can be chosen as
suited for a specific situation. Hence the adapted metric using (\ref{eq:grrT})
and (\ref{eq:gththT}) becomes
\begin{equation}
ds^{2}=-N(T)^{2}dT^{2}+\frac{p_{b}^{2}}{L_{0}^{2}|p_{c}|}dx^{2}+|p_{c}|(d\theta^{2}+\sin^{2}\theta d\phi^{2})
\end{equation}
Comparing this metric written in Schwarzschild coordinates and lapse
$N(t)$, with the standard Schwarzschild interior metric but with
rescaled $r\to lx$
\begin{equation}
ds^{2}=-\left(\frac{2GM}{t}-1\right)^{-1}dt^{2}+l^{2}\left(\frac{2GM}{t}-1\right)dx^{2}+t^{2}\left(d\theta^{2}+\sin^{2}\theta d\phi^{2}\right),
\end{equation}
we see that 
\begin{align}
N\left(t\right)= & \left(\frac{2GM}{t}-1\right)^{-\frac{1}{2}},\label{eq:Sch-corresp-1}\\
g_{xx}\left(t\right)= & \frac{p_{b}\left(t\right)^{2}}{L_{0}^{2}|p_{c}\left(t\right)|}=l^{2}\left(\frac{2GM}{t}-1\right),\label{eq:Sch-corresp-2}\\
g_{\theta\theta}\left(T\right)= & \frac{g_{\phi\phi}\left(T\right)}{\sin^{2}\left(\theta\right)}=g_{\Omega\Omega}\left(T\right)=|p_{c}\left(t\right)|=t^{2}.\label{eq:Sch-corresp-3}
\end{align}
This shows that 
\begin{align}
p_{b}= & 0, & p_{c}= & 4G^{2}M^{2}, &  & \textrm{On the horizon\,}t=2GM,\label{eq:t-horiz}\\
p_{b}\to & 0, & p_{c}\to & 0, &  & \textrm{At the singularity\,}t\to0.\label{eq:t-singular}
\end{align}
Also, note that in the fiducial volume, we can consider three surfaces
$S_{x,\theta},\,S_{x,\phi}$ and $S_{\theta,\phi}$ respectively bounded
by $\mathcal{I}$ and a great circle along a longitude of $V_{0}$,
$\mathcal{I}$ and the equator of $V_{0}$, and the equator and a
longitude with areas \cite{Ashtekar:2005qt}
\begin{align}
A_{x,\theta}=A_{x,\phi}= & 2\pi L_{0}\sqrt{g_{xx}g_{\Omega\Omega}}=2\pi p_{b}\label{eq:area-x}\\
A_{\theta,\phi}= & \pi g_{\Omega\Omega}=\pi |p_{c}|\label{eq:area-angl}
\end{align}
with the volume of the fiducial region $\mathcal{I}\times\mathbb{S}^{2}$
given by \cite{Ashtekar:2005qt}
\begin{equation}
V=\int\mathrm{d}^{3}x\sqrt{|\det\tilde{E}|}=4\pi L_0\sqrt{g_{xx}}g_{\Omega\Omega}=4\pi p_{b}\sqrt{|p_{c}|},
\end{equation}
where $\sqrt{\det|\tilde{E}|}=\sqrt{q}$ with $q$ being the determinant
of the spatial metric. It is also worth noting that the Riemann invariants
such the Kretschmann scalar which in this case is
\begin{equation}
R_{abcd}R^{abcd}\propto\frac{1}{p_{c}^{3}},\label{eq:Kretsch}
\end{equation}
diverge at the singularity as expected.

\section{Classical dynamics\label{sec:Classical-dynamics}}

In order to be able to compare the effects of GUP, we first need to
have the classical dynamics. Choosing a lapse
\begin{equation}
N\left(T\right)=\frac{\gamma\, \mathrm{sgn}(p_c)\sqrt{|p_{c}\left(T\right)|}}{b\left(T\right)}\label{eq:lapsNT}
\end{equation}
the Hamiltonian constraint (\ref{eq:H-class-N}) becomes
\begin{equation}
H=-\frac{1}{2G\gamma}\left[\left(b^{2}+\gamma^{2}\right)\frac{p_{b}}{b}+2cp_{c}\right].\label{eq:H-class-1}
\end{equation}
The reason for choosing the lapse (\ref{eq:lapsNT}) is that as we
will see, the equations of motion of the pair $(c,p_{c})$ decouple
from those of $(b,p_{b})$. 

As far as the classical physical results go, choice of a lapse does
not matter and equations of motion are equivalent. We will extend
our study in a future work to a corresponding quantum model in which
we will use a different lapse, but clearly the classical counterpart
of such a different lapse is equivalent to the above choice of lapse.

The equations of motion corresponding to (\ref{eq:H-class-1}) are
\begin{align}
\frac{db}{dT}= & \left\{ b,H\right\} =-\frac{1}{2}\left(b+\frac{\gamma^{2}}{b}\right),\label{eq:EoM-diff-b}\\
\frac{dp_{b}}{dT}= & \left\{ p_{b},H\right\} =\frac{p_{b}}{2}\left(1-\frac{\gamma^{2}}{b^{2}}\right).\label{eq:EoM-diff-pb}\\
\frac{dc}{dT}= & \left\{ c,H\right\} =-2c,\label{eq:EoM-diff-c}\\
\frac{dp_{c}}{dT}= & \left\{ p_{c},H\right\} =2p_{c},\label{eq:EoM-diff-pc}
\end{align}
These equations should be supplemented by the weakly vanishing ($\approx0$)
of the Hamiltonian constraint (\ref{eq:H-class-1}),
\begin{equation}
\left(b^{2}+\gamma^{2}\right)\frac{p_{b}}{b}+2cp_{c}\approx0.\label{eq:weak-van}
\end{equation}
Before solving these equations, let us clarify the interpretation
of these variables. It is clear from (\ref{eq:Sch-corresp-3}) that
$p_{c}$ is the square of the radius of the infalling 2-spheres. The
interpretation of $p_{b}$ is also clear from (\ref{eq:area-x}).
In order to better understand the role of $b,\,c$ we use the relation
of the proper time $\tau$ and a generic time $T$ for the metric
(\ref{eq:K-S-gener}),
\begin{equation}
d\tau^{2}=-N(T)^{2}dT^{2},
\end{equation}
and the form of the lapse function (\ref{eq:lapsNT}), to rewrite
equations (\ref{eq:EoM-diff-pc}) as
\begin{equation}
b=\frac{\gamma}{2}\frac{1}{\sqrt{p_{c}}}\frac{dp_{c}}{d\tau}=\gamma\frac{d}{d\tau}\sqrt{g_{\Omega\Omega}}=\frac{\gamma}{\sqrt{\pi}}\frac{d}{d\tau}\sqrt{A_{\theta,\phi}},\label{eq:b-interp}
\end{equation}
where the last two terms on the right hand side were derived using
(\ref{eq:area-angl}). Hence, classically, $b$ is proportional to
the rate of change of the square root of the physical area of $\mathbb{S}^{2}$. 

To interpret the role of $c$, we combine (\ref{eq:EoM-diff-pb}),
(\ref{eq:b-interp}), (\ref{eq:weak-van}), and (\ref{eq:lapsNT})
together with (\ref{eq:area-x}) and (\ref{eq:area-angl}), to get
\begin{equation}
c=\gamma\frac{d}{d\tau}\left(\frac{p_{b}}{\sqrt{p_{c}}}\right)=\gamma\frac{d}{d\tau}\left(L_{0}\sqrt{g_{xx}}\right).
\end{equation}
Hence, classically $c$ is proportional to the rate of change of the
physical length of $\mathcal{I}$.

The solution to the system of differential equations (\ref{eq:EoM-diff-b})-(\ref{eq:EoM-diff-pc})
are
\begin{align}
b\left(T\right)= & \pm\sqrt{e^{2C_{1}}e^{-T}-\gamma^{2}},\\
p_{b}\left(T\right)= & C_{2}e^{\frac{T}{2}}\sqrt{e^{2C_{1}}-\gamma^{2}e^{T}},\\
c\left(T\right)= & C_{3}e^{-2T},\\
p_{c}\left(T\right)= & C_{4}e^{2T}.\label{eq:EoM-Sol-T-pc}
\end{align}
In order to find the constants of integration, we note that if we
write the solutions in Schwarzschild time $t$, then from (\ref{eq:Sch-corresp-3})
we should have $p_{c}(t)=t^{2}$. Having this in mind and looking
at (\ref{eq:EoM-Sol-T-pc}), we see that a transformation 
\begin{equation}
T=\ln\left(t\right)
\end{equation}
 will give us a similar form. Using such a transformation we get
\begin{align}
b\left(t\right)= & \pm\sqrt{\frac{e^{2C_{1}}}{t}-\gamma^{2}},\label{eq:b-t-class-sol}\\
p_{b}\left(t\right)= & C_{2}t\sqrt{\frac{e^{2C_{1}}}{t}-\gamma^{2}},\label{eq:pb-t-class-sol}\\
c\left(t\right)= & \frac{C_{3}}{t^{2}},\label{eq:c-t-class-sol}\\
p_{c}\left(t\right)= & C_{4}t^{2}.\label{eq:pc-t-class-sol}
\end{align}
Comparing with $p_{c}(t)=t^{2}$ we see that $C_{4}=1$. Also from
(\ref{eq:t-horiz}) we can deduce
\begin{equation}
0=p_{b}\left(2GM\right)=2GMC_{2}\sqrt{\frac{e^{2C_{1}}}{2GM}-\gamma^{2}},
\end{equation}
which yields
\begin{equation}
C_{1}=\frac{1}{2}\ln\left(2GM\gamma^{2}\right).\label{eq:C1-EoM}
\end{equation}
Next, we see from (\ref{eq:Sch-corresp-2}) that
\begin{equation}
p_{b}\left(t\right)^{2}=l^{2}\left(\frac{2GM}{t}-1\right)L_{0}^{2}t^{2}
\end{equation}
which if compared with (\ref{eq:pb-t-class-sol}) and using (\ref{eq:C1-EoM})
yields
\begin{equation}
C_{2}=\frac{lL_{0}}{\gamma}.\label{eq:C2-EoM}
\end{equation}
Finally using (\ref{eq:weak-van}), we get 
\begin{equation}
C_{3}=\mp\gamma GMlL_{0}.\label{eq:C3-EoM}
\end{equation}
Thus, the solutions to the equations of motion in terms of Schwarzschild
time $t$ become
\begin{align}
b\left(t\right)= & \pm\gamma\sqrt{\frac{2GM}{t}-1},\label{eq:sol-cls-b}\\
p_{b}\left(t\right)= & lL_{0}t\sqrt{\frac{2GM}{t}-1},\label{eq:sol-cls-pb}\\
c\left(t\right)= & \mp\frac{\gamma GMlL_{0}}{t^{2}},\label{eq:sol-cls-c}\\
p_{c}\left(t\right)= & t^{2}.\label{eq:sol-cls-pc}
\end{align}

\begin{figure}
\begin{centering}
\includegraphics[scale=0.7]{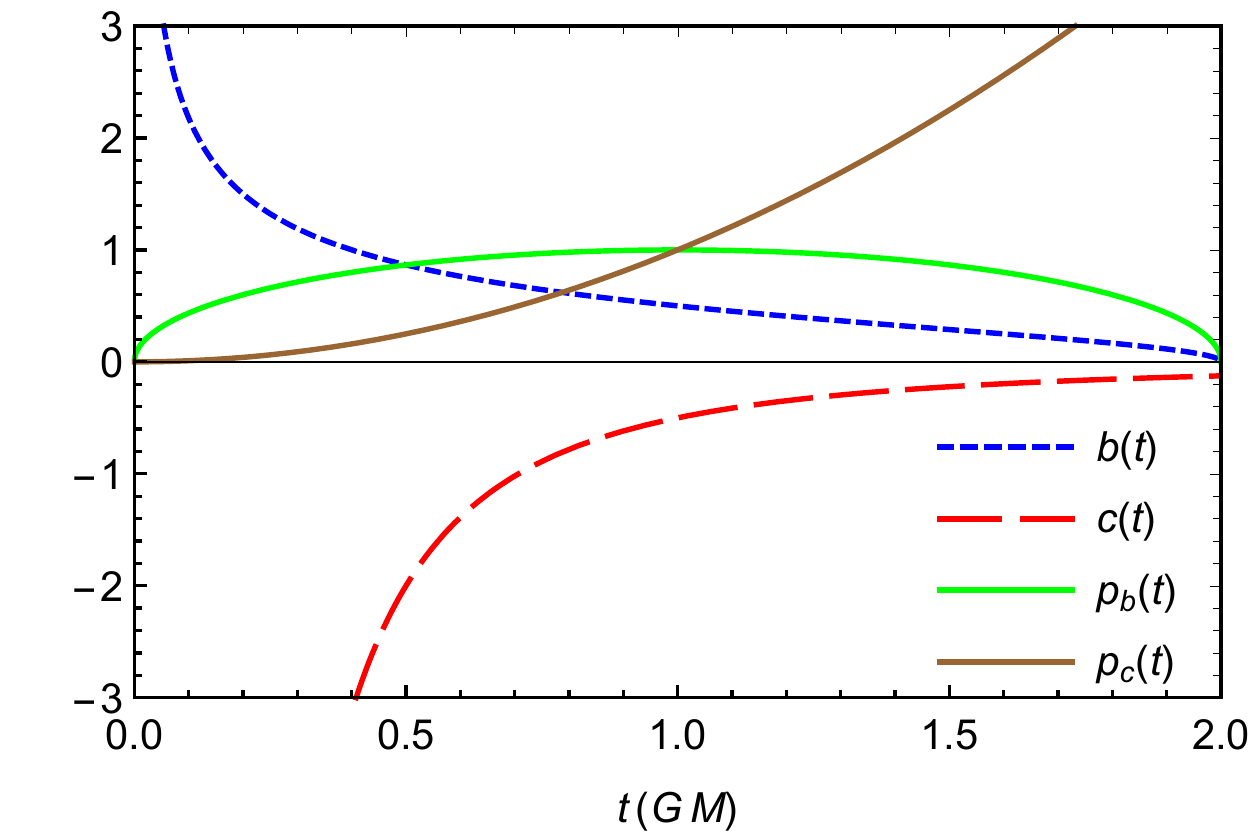}
\par\end{centering}
\caption{The behavior of canonical variables as a function of Schwarzschild
time $t$. We have chosen the positive sign for $b$ and negative
sign for $c$. The figure is plotted using $\gamma=0.5,\,M=1,\,G=1$
and $L_{0}=1$. \label{fig:class-var-behv}}
\end{figure}

The behavior of these solutions as a function of $t$ is depicted
in Fig. \ref{fig:class-var-behv}. From these equations or the plot,
one can see that $p_{c}\to0$ as $t\to0$, i.e., at the classical
singularity, leading to the Riemann invariants such as the Kretschmann
scalar
\begin{equation}
K=R_{abcd}R^{abcd}\propto\frac{1}{p_{c}^{3}},
\end{equation}
all diverge, signaling the presence of a physical singularity there
as expected.

We can also see from Fig. \ref{fig:class-var-behv} that $b$, the
rate of change of the square root of the physical area of $\mathbb{S}^{2}$,
as well as $c$, the rate of change of the physical length of $\mathcal{I}$,
diverge at the classical singularity.

\section{Effective dynamics inspired by GUP\label{sec:Effective-GUP-dynamics}}

\subsection{Deformation of Poisson brackets\label{subsec:GUP-redef-PB}}

In order to find the effective GUP-modified dynamics, we impose a minimal uncertainty in $p_{b}$ and $p_{c}$, and thus we modify the classical algebra of variables.
To be as general as possible, let
us call the configuration variables $q_{1},\,q_{2}$ and the momenta
$p_{1},\,p_{2}$. In our case
\begin{align}
q_{1}= & b, & q_{2}= & c,\label{eq:bcqq}\\
p_{1}= & \frac{1}{G\gamma}p_{b}, & p_{2}= & \frac{1}{2G\gamma}p_{c}.\label{eq:pbpcpp}
\end{align}
Our purpose here is to impose alternative relations to (\ref{eq:classic-PBs-bc})
in order to reproduce GUP effects. For this reason, it is convenient
to define a new pair of configuration variables conjugate to $p_{i}$.
That is, we introduce the quantities $\bar{q}_{1}$ and
$\bar{q}_{2}$ such that 
\begin{align}
\left\{ \bar{q}_{1},p_{1}\right\} = & 1,\\
\left\{ \bar{q}_{2},p_{2}\right\} = & 1.
\end{align}
Thus, we modify the algebra such that
\begin{align}
\left\{ q_{1},p_{1}\right\} _{\bar{q},p}= & f\left(q_{1},q_{2}\right),\label{eq:PB-f-1}\\
\left\{ q_{2},p_{2}\right\} _{\bar{q},p}= & g\left(q_{1},q_{2}\right),\label{eq:PB-g-1}
\end{align}
with the rest of Poisson brackets vanishing.
This means that pairs $(q_{1},p_{1})$, as well as $(q_{2},p_{2})$, are no longer canonically conjugate.
Note that the Poisson brackets are evaluated with respect
to $\bar{q}_{i},p_{i}$.
The quantities $\bar{q}_{i}$ can then be
constructed starting from (\ref{eq:PB-f-1}) and (\ref{eq:PB-g-1}), that is 
\begin{align}
\frac{\partial q_{1}}{\partial\bar{q}_{1}}= & f\left(q_{1},p_{1}\right),\\
\frac{\partial q_{2}}{\partial\bar{q}_{2}}= & g\left(q_{2},p_{2}\right),
\end{align}
whence
\begin{align}
    \bar{q}_{1}= & \int_{q_{1(0)}}^{q_{1}}\frac{dq_{1}^{\prime}}{f\left(q_{1}^{\prime},p_{1}\right)}, \label{eq:q1-relat-q1}\\
    \bar{q}_{2}= & \int_{q_{2(0)}}^{q_{2}}\frac{dq_{2}^{\prime}}{g\left(q_{2}^{\prime},p_{2}\right).}\label{eq:q2-relat-q2}
\end{align}
In this work, we consider functions $f,\,g$ such that
\begin{align}
f\left(q_{1},q_{2}\right)= & f\left(q_{1}\right)=1+\beta_{1} q_{1}^{2},\\
g\left(q_{1},q_{2}\right)= & g\left(q_{2}\right)=1+\beta_{2} q_{2}^{2}.
\end{align}
where $\beta_{1}$ and $\beta_{2}$ are suitable dimensional parameters.
In other words, we assume that any effects due to a minimal
uncertainty in $p_{1}$ does not influence $p_{2}$ and vice versa.
Such choice corresponds to particular deformations of the canonical algebra in (\ref{eq:PB-f-1}) and (\ref{eq:PB-g-1}) originating from GUP.
In fact, considering the corresponding GUP commutation relations in a quantum description
and computing from them the uncertainty relations, we would find that $\Delta p_1$ and $\Delta p_2$ are bounded from below.
This effectively describes minimal uncertainties in $p_1$ and $p_2$.
Such minimal uncertainties are proportional to $\sqrt{|\beta_1|}$ and $\sqrt{|\beta_2|}$.
The cases $\beta_1,\beta_2<0$, suggested by some works (see, \emph{e.g.}, \cite{Jizba:2009qf,Ong:2018zqn,Buoninfante:2019fwr}), deserve a dedicated comment.
In fact, negative values of $\beta_1$ and $\beta_2$ imply that the functions $f(q_1)$ and $g(q_2)$ may acquire negative values.
However, this is not allowed for $f(q_1)$ and $g(q_2)$ are ultimately related, in this model, to the measure of the space described by $q_1$ and $q_2$.
The regions in which the functions $f(q_1)$ and $g(q_2)$ become negative can be excluded by restricting the intervals on which the quantities $q_1$ and $q_2$ acquire value.
Specifically, this imposes maximum values for $|q_1|$ and $|q_2|$.
As we will see, such a feature is compatible with Eqs. \eqref{eq:q1-relat-q1} and \eqref{eq:q2-relat-q2} below.
Furthermore, as shown in \cite{Bosso:2020aqm}, the auxiliary operators $\bar{q}_1$ and $\bar{q}_2$, which are bounded for $\beta_1,~\beta_2>0$, for negative parameters become unbounded and are represented by the usual operators $\bar{q}_1 = i \frac{d}{d p_1}$ and $\bar{q}_2 = i \frac{d}{dp_2}$.
Thus, such operators are symmetric.
Moreover, due to Eqs.\eqref{eq:q1-relat-q1} and \eqref{eq:q2-relat-q2}, this ensures the operators $q_1$ and $q_2$ are symmetric as well.
We will consider such aspects related to the quantum version of the model presented here in a future work \cite{preparation}.

\subsection{Modified dynamics \label{subsec:GUP-modified-dynamics}}

Using the specific variables of our model $\left(b,p_{b},c,p_{c}\right)$,
the algebra (\ref{eq:PB-f-1})-(\ref{eq:PB-g-1}) becomes
\begin{align}
\left\{ b,p_{b}\right\} = & G\gamma\left(1+\beta_{b}b^{2}\right),\label{eqn:b_pb}\\
\left\{ c,p_{c}\right\} = & 2G\gamma\left(1+\beta_{c}c^{2}\right),\label{eqn:c_pc}
\end{align}
where we have renamed $\beta_{1}\to\beta_{b}$ and $\beta_{2}\to\beta_{c}$.
It is worth emphasize that, as done in the previous section, because of such modifications the pairs of variables $(b,p_b)$ and $(c,p_c)$ do not constitute pairs of canonically conjugate variables.
In fact, configuration variables canonically conjugate to $p_b$ and $p_c$ would be two auxiliary variables $\bar{b}$ and $\bar{c}$, respectively, defined in Eqs. \eqref{eq:q1-relat-q1} and \eqref{eq:q2-relat-q2} \cite{Bosso:2018uus,Bosso:2020aqm}.
When the modified algebra above is regarded in a quantum context,
it implies a minimal uncertainty in $p_{b}$ and $p_{c}$ \cite{Kempf:1994su}.
In fact, considering the corresponding commutation relations 
\begin{align}
\left[b,p_{b}\right]= & iG\gamma\left(1+\beta_{b}b^{2}\right),\\
\left[c,p_{c}\right]= & i2G\gamma\left(1+\beta_{c}c^{2}\right),
\end{align}
one can find the following uncertainty relations 
\begin{align}
\Delta b\Delta p_{b}\geq & \frac{G\gamma}{2}\left[1+\beta_{b}(\Delta b)^{2}\right],\label{eqn:unc_b_pb}\\
\Delta c\Delta p_{c}\geq & G\gamma\left[1+\beta_{c}(\Delta c)^{2}\right],\label{eqn:unc_c_pc}
\end{align}
which correspond to minimal uncertainties for $p_{b}$ and $p_{c}$
of the order of $G\gamma\sqrt{\beta_{b}}$ and $2G\gamma\sqrt{\beta_{c}}$,
respectively. Therefore, $\beta_{b}$ and $\beta_{c}$ effectively
define the magnitude of the effects introduced with the algebra (\ref{eqn:b_pb})-(\ref{eqn:c_pc}).

Using this new algebra and the Hamiltonian (\ref{eq:H-class-1}),
the new GUP-modified equations of motion become
\begin{align}
\frac{db}{dT}= & \left\{ b,H\right\} =-\frac{1}{2}\left(b+\frac{\gamma^{2}}{b}\right)\left(1+\beta_{b}b^{2}\right),\label{eq:EoM-diff-b-1}\\
\frac{dp_{b}}{dT}= & \left\{ p_{b},H\right\} =\frac{p_{b}}{2}\left(1-\frac{\gamma^{2}}{b^{2}}\right)\left(1+\beta_{b}b^{2}\right),\label{eq:EoM-diff-pb-1}\\
\frac{dc}{dT}= & \left\{ c,H\right\} =-2c\left(1+\beta_{c}c^{2}\right),\label{eq:EoM-diff-c-1}\\
\frac{dp_{c}}{dT}= & \left\{ b,H\right\} =2p_{c}\left(1+\beta_{c}c^{2}\right).\label{eq:EoM-diff-pc-1}
\end{align}
The solution to such modified equations of motion are
\begin{align}
b\left(T\right)= & \pm\frac{\sqrt{e^{2C_{1}}e^{\beta_{b}\gamma^{2}T}-\gamma^{2}e^{2\beta_{b}\gamma^{2}C_{1}}e^{T}}}{\sqrt{e^{2\beta_{b}\gamma^{2}C_{1}}e^{T}-\beta_{b}e^{2C_{1}}e^{\beta_{b}\gamma^{2}T}}},\\
p_{b}\left(T\right)= & C_{2}e^{-\beta_{b}\gamma^{2}T}\sqrt{\left(e^{2\beta_{b}\gamma^{2}C_{1}}e^{T}-\beta_{b}e^{2C_{1}}e^{\beta_{b}\gamma^{2}T}\right)\left(e^{2C_{1}}e^{\beta_{b}\gamma^{2}T}-\gamma^{2}e^{2\beta_{b}\gamma^{2}C_{1}}e^{T}\right)},\\
c\left(T\right)= & \mp\frac{e^{C_{3}}}{\sqrt{e^{4T}-\beta_{c}e^{2C_{3}}}},\\
p_{c}\left(T\right)= & C_{4}\sqrt{e^{4T}-\beta_{c}e^{2C_{3}}}.
\end{align}
where $C_{i}$, with $i=1,\cdots,4$ are integration constants.
Once again, we make a transformation to the Schwarzschild time $t$ using
$T=\ln(t)$. This leads to
\begin{align}
b\left(t\right)= & \pm\frac{\sqrt{e^{2C_{1}}t^{\beta_{b}\gamma^{2}}-\gamma^{2}te^{2\beta_{b}\gamma^{2}C_{1}}}}{\sqrt{te^{2\beta_{b}\gamma^{2}C_{1}}-\beta_{b}e^{2C_{1}}t^{\beta_{b}\gamma^{2}}}},\\
p_{b}\left(t\right)= & C_{2}t^{-\beta_{b}\gamma^{2}}\sqrt{\left(te^{2\beta_{b}\gamma^{2}C_{1}}-\beta_{b}e^{2C_{1}}t^{\beta_{b}\gamma^{2}}\right)\left(e^{2C_{1}}t^{\beta_{b}\gamma^{2}}-\gamma^{2}te^{2\beta_{b}\gamma^{2}C_{1}}\right)},\\
c\left(t\right)= & \mp\frac{e^{C_{3}}}{\sqrt{t^{4}-\beta_{c}e^{2C_{3}}}},\\
p_{c}\left(t\right)= & C_{4}\sqrt{t^{4}-\beta_{c}e^{2C_{3}}}.
\end{align}

\begin{figure}
\begin{centering}
\includegraphics[scale=0.57]{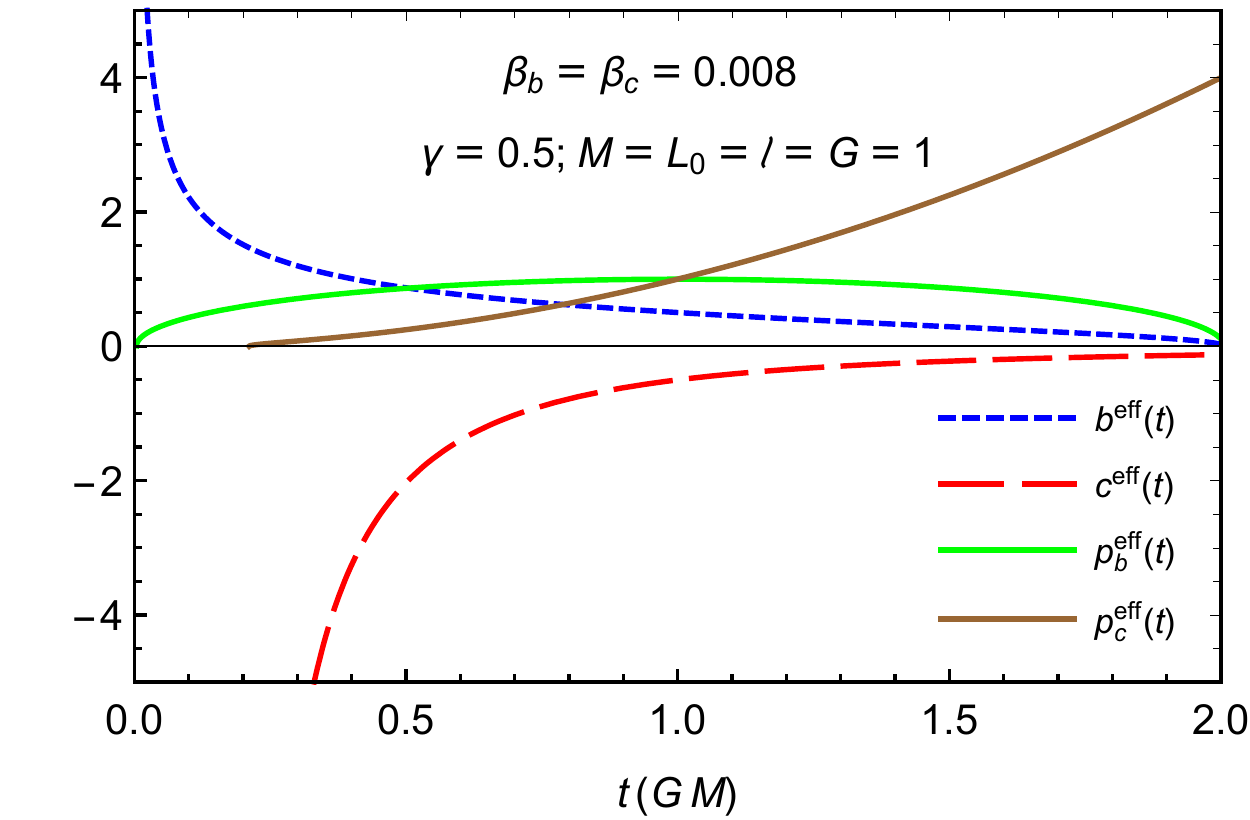}~~~~\includegraphics[scale=0.57]{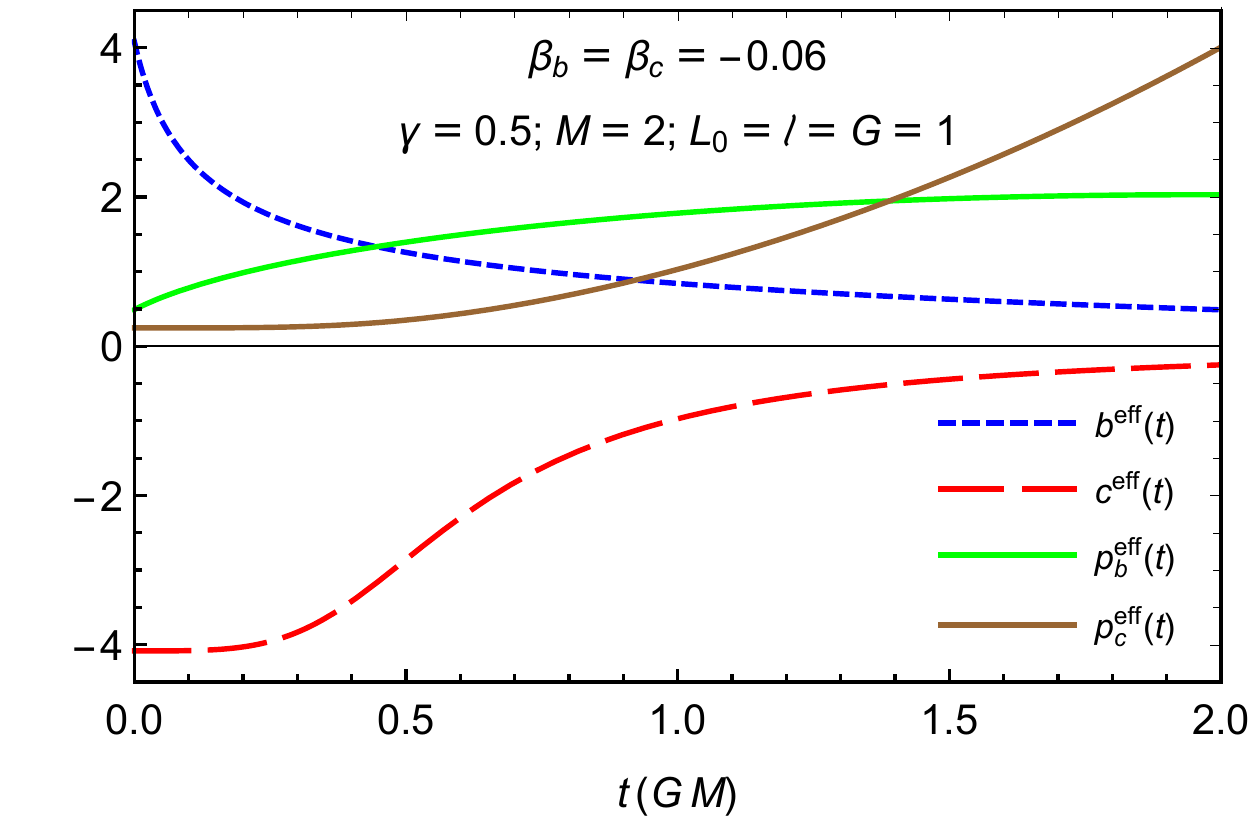}
\par\end{centering}
\begin{centering}
\vspace{10pt}
\includegraphics[scale=0.57]{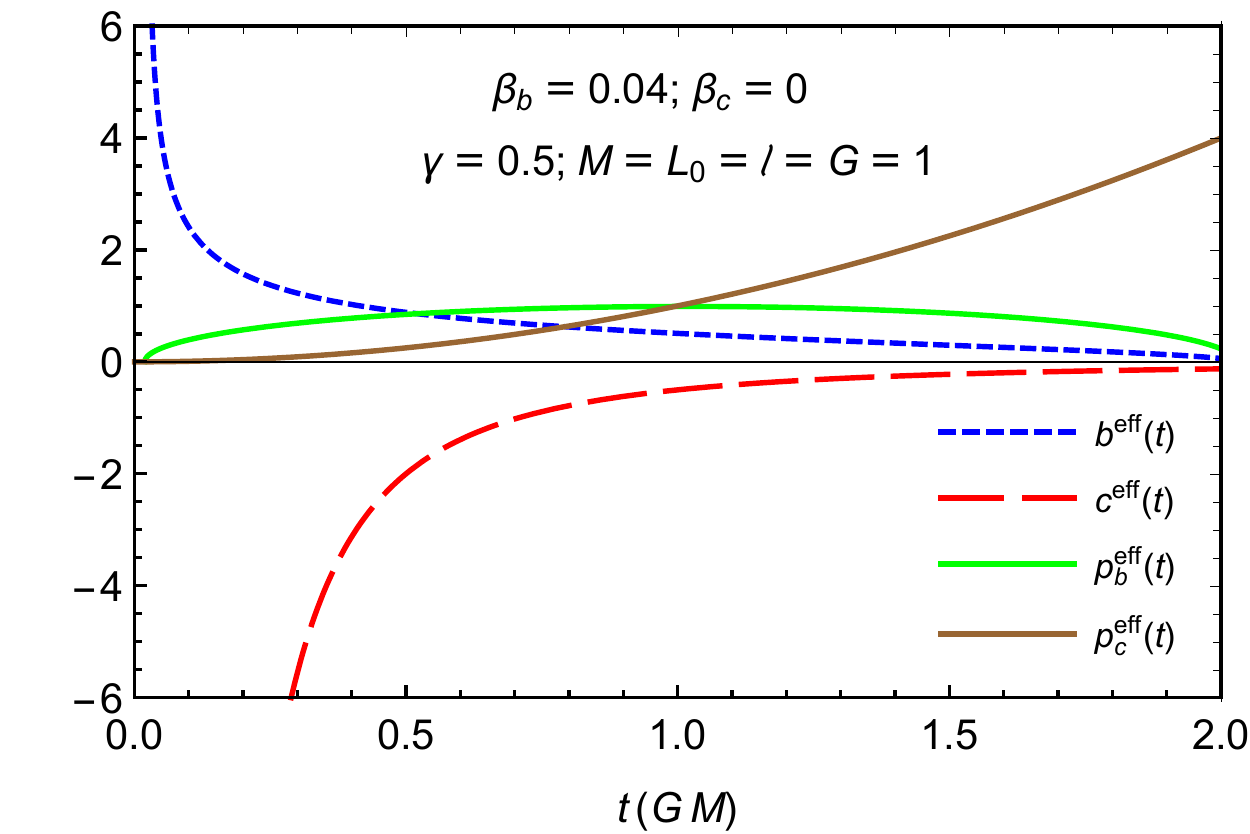}~~~~\includegraphics[scale=0.57]{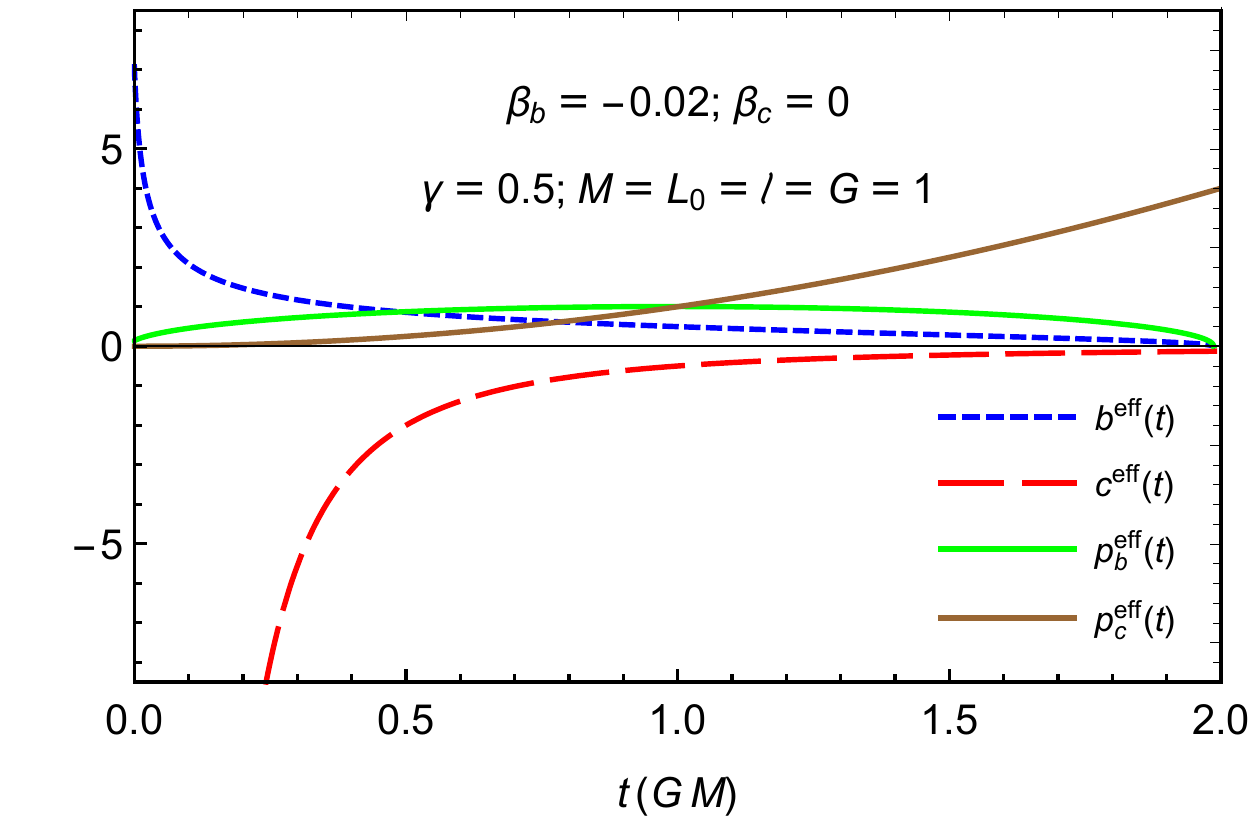}
\par\end{centering}
\begin{centering}
\vspace{10pt}
\includegraphics[scale=0.57]{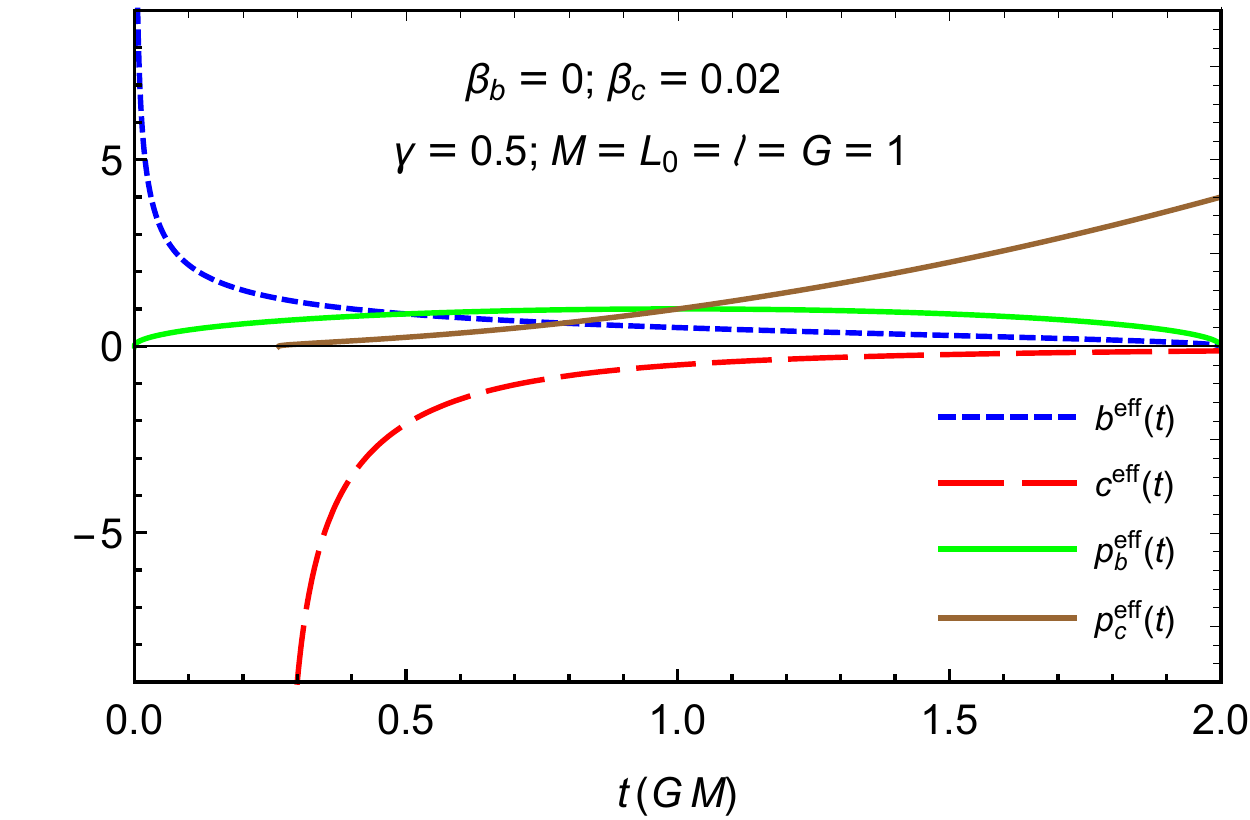}~~~~\includegraphics[scale=0.57]{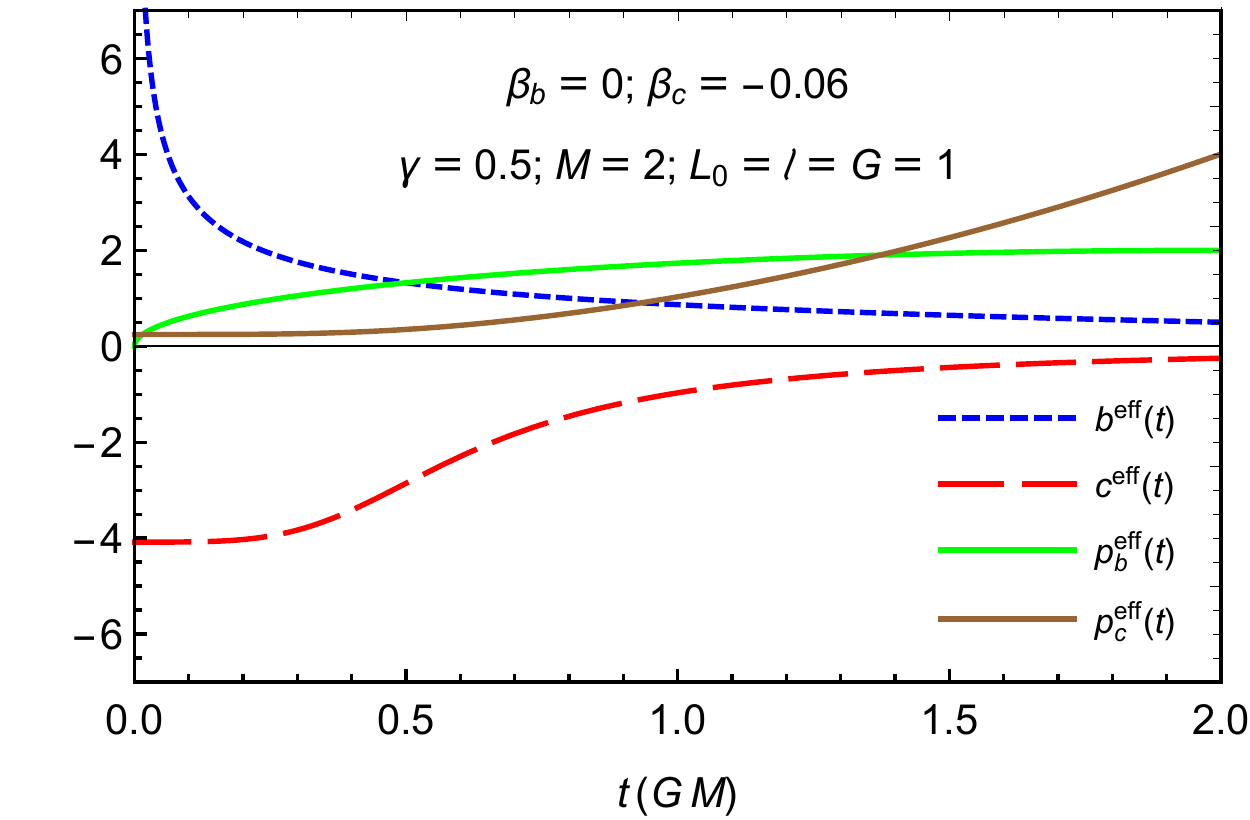}
\par\end{centering}
\caption{The behavior of solutions of the modified case in Schwarzschild time
$t$ for positive, negative and vanishing $\beta_{b}$ and $\beta_{c}$ for the whole interior.
We have chosen the positive sign for $b$ and negative sign for $c$. Note that for nonvanishing negative $\beta_c$ we always get a minimum nonvanishing value for $p_c$, while a nonvanishing negative $\beta_b$ leads to a finite value of $b$ at $t\to 0$.
The values of parameters are mentioned on each plot.
\label{fig:GUP-1-pos-neg}}
\end{figure}

We can find the value of the integration constant by considering the standard limit, \emph{i.e.}, the limit with $\beta_{b},\beta_{c}\to0$. In such limit, requiring $p_{c}(t)$ to have the form (\ref{eq:Sch-corresp-3}),
we obtain $C_{4}=1$. Furthermore, since both $b(t)$ and $p_{b}(t)$
acquire the corresponding classical form (\ref{eq:b-t-class-sol})
and (\ref{eq:pb-t-class-sol}), respectively, the corresponding integration
constants are the same as in (\ref{eq:C1-EoM}) and (\ref{eq:C2-EoM}).
Finally, since the classical limit of $c(t)$ is $\mp\frac{e^{C_{3}}}{t^{2}}$,
comparing it with (\ref{eq:c-t-class-sol}) and (\ref{eq:C3-EoM}),
we obtain 
\begin{equation}
e^{C_{3}}=\gamma GMlL_{0}.
\end{equation}
Substituting the integration constants back into the equations of
motion yields 
\begin{align}
b\left(t\right)= & \pm\frac{\gamma\sqrt{2GMt^{\beta_{b}\gamma^{2}}-t\left(2\gamma^{2}GM\right)^{\beta_{b}\gamma^{2}}}}{\sqrt{t\left(2\gamma^{2}GM\right)^{\beta_{b}\gamma^{2}}-2\beta_{b}\gamma^{2}GMt^{\beta_{b}\gamma^{2}}}},\label{eq:b-eff-t}\\
p_{b}\left(t\right)= & lL_{0}t^{-\beta_{b}\gamma^{2}}\sqrt{\left[2GMt^{\beta_{b}\gamma^{2}}-t\left(2\gamma^{2}GM\right)^{\beta_{b}\gamma^{2}}\right]\left[t\left(2\gamma^{2}GM\right)^{\beta_{b}\gamma^{2}}-2\beta_{b}\gamma^{2}GMt^{\beta_{b}\gamma^{2}}\right]},\\
c\left(t\right)= & \mp\frac{\gamma GMlL_{0}}{\sqrt{t^{4}-\beta_{c}\gamma^{2}G^{2}M^{2}l^{2}L_{0}^{2}}},\label{eq:c-eff-t}\\
p_{c}\left(t\right)= & \sqrt{t^{4}-\beta_{c}\gamma^{2}G^{2}M^{2}l^{2}L_{0}^{2}}.\label{eq:pc-eff-t}
\end{align}
The behavior of these solutions is depicted in Fig. \ref{fig:GUP-1-pos-neg}.
In general, it is seen that positive values of $\beta_{b}$ and/or $\beta_{c}$ (left figures) either do not affect the classical behavior of $p_{c}$, hence not resolving the singularity, or lead to the abrupt interruption of the evolution of $p_{c}$ close to the classical singularity.
On the other hand, in both cases in which $\beta_{b},\beta_{c}<0$
(top right figure) and $\beta_{b}=0,\,\beta_{c}<0$ (bottom right
figure), the effective dynamics of $p_{c}$ is modified such that
it exhibits a minimum value.
Hence, the Riemann invariant does not blow up anywhere inside the black hole.
The requirement for the parameters $\beta_{b}$
and $\beta_{c}$ to be negative is consistent with previous and independent
analyses (see, \emph{e.g.}, \cite{Ong:2018zqn}).
Furthermore, it is worth noticing that in the above solutions any instance of $\beta_{i}$ is accompanied by a factor $\gamma^{2}$, while the inverse does not hold. 

\subsubsection{A prescription to remove the dependence on the fiducial $L_{0}$  \label{subsec:depend_L0} }

It is seen that the solutions \eqref{eq:b-eff-t}-\eqref{eq:pc-eff-t},
particularly $p_{c}$, depend on the fiducial length $L_{0}$. This
is an issue which usually appears in the quantization of the interior
of the Schwarzschild black hole as a classical symmetry reduced model
\cite{Ashtekar:2005qt}. In LQG, this is usually cured by making the
polymer parameters of the theory either momentum dependent \cite{Bohmer:2007wi,Chiou:2008nm}
($\bar{\mu}$ schemes or improved dynamics), or mass dependent \cite{Corichi:2015xia}. 

In this work we propose a prescription that is closer to the ideas
in \cite{Corichi:2015xia}. We define
\begin{equation}
-\beta_{c}L_{0}^{2}=\ell_{c}^{2}\label{eq:prescript}
\end{equation}
where $\ell_{c}$ is a fundamental physical minimum length scale of
the theory. This is inspired by the fact that the factor $-\beta_{c}L_{0}^{2}$
appears in both \eqref{eq:c-eff-t} and \eqref{eq:pc-eff-t}. We will
also see that such a definition makes expansion and shear scalars
independent of $L_{0}$ and furthermore leads to an interesting relation
between LQG and GUP minimum scales. 

Following this prescription \eqref{eq:prescript}, the equations of
motion \eqref{eq:b-eff-t}-\eqref{eq:pc-eff-t} can now be rewritten
as (with $l=1$)
\begin{align}
b\left(t\right)= & \pm\frac{\gamma\sqrt{2GMt^{\beta_{b}\gamma^{2}}-t\left(2\gamma^{2}GM\right)^{\beta_{b}\gamma^{2}}}}{\sqrt{t\left(2\gamma^{2}GM\right)^{\beta_{b}\gamma^{2}}-2\beta_{b}\gamma^{2}GMt^{\beta_{b}\gamma^{2}}}},\label{eq:b-eff-t-no-L0}\\
p_{b}\left(t\right)= & \frac{\ell_{c}}{\sqrt{-\beta_{c}}}t^{-\beta_{b}\gamma^{2}}\sqrt{\left[2GMt^{\beta_{b}\gamma^{2}}-t\left(2\gamma^{2}GM\right)^{\beta_{b}\gamma^{2}}\right]\left[t\left(2\gamma^{2}GM\right)^{\beta_{b}\gamma^{2}}-2\beta_{b}\gamma^{2}GMt^{\beta_{b}\gamma^{2}}\right]},\label{eq:pb-eff-t-no-L0}\\
c\left(t\right)= & \mp\frac{\ell_{c}}{\sqrt{-\beta_{c}}}\frac{\gamma GM}{\sqrt{t^{4}+\ell_{c}^{2}\gamma^{2}G^{2}M^{2}}},\label{eq:c-eff-t-no-L0}\\
p_{c}\left(t\right)= & \sqrt{t^{4}+\ell_{c}^{2}\gamma^{2}G^{2}M^{2}}.\label{eq:pc-eff-t-no-L0}
\end{align}
It is clear that the quantity $p_{c}$ which represents the radius
of the infalling 2-spheres does not depend on $L_{0}$ anymore.
Furthermore, both the expansion
\begin{align}
\theta= & \frac{\dot{p}_{b}}{Np_{b}}+\frac{\dot{p}_{c}}{2Np_{c}}\nonumber \\
= & \frac{1}{2\gamma\sqrt{p_{c}}}\left[\beta_{b}b\left(b^{2}-\gamma^{2}\right)+2b\beta_{c}c^{2}+3b-\frac{\gamma^{2}}{b}\right]
\end{align}
and the shear scalar
\begin{align}
\sigma^{2}= & \frac{2}{3}\left(\frac{\dot{p}_{b}}{Np_{b}}-\frac{\dot{p}_{c}}{Np_{c}}\right)^{2}\nonumber \\
= & \frac{2}{p_{c}}\left[\beta_{b}\left(-\frac{b^{4}}{2\gamma^{2}}+\frac{b^{2}}{3}+\frac{\gamma^{2}}{6}\right)+\beta_{c}c^{2}\left(\frac{2b^{2}}{\gamma^{2}}+\frac{2}{3}\right)+\beta_{b}^{2}\left(\frac{b^{6}}{12\gamma^{2}}-\frac{b^{4}}{6}+\frac{b^{2}\gamma^{2}}{12}\right)\right.\nonumber \\
 & \left.+\frac{4b^{2}\beta_{c}^{2}c^{4}}{3\gamma^{2}}+\beta_{b}\beta_{c}c^{2}\left(\frac{2b^{2}}{3}-\frac{2b^{4}}{3\gamma^{2}}\right)+\frac{3b^{2}}{4\gamma^{2}}+\frac{\gamma^{2}}{12b^{2}}+\frac{1}{2}\right]
\end{align}
are independent of $L_{0}$, since $b$ is independent of $L_{0}$,
$p_{b}$ does not appear in either expressions, and every instance of
$c$ is accompanied by a $\beta_{c}^{2}$. The combination $\beta_{c}c^{2}$
is independent of $L_{0}$ because as seen from \eqref{eq:c-eff-t-no-L0},
$c^{2}$ comes with a factor $\beta_{c}^{-1}$. Additionally, the
right hand side of \eqref{eqn:c_pc} remains independent of the fiducial
length $L_{0}$. 

\subsubsection{Minimum value of $p_{c}$, comparison to LQG, and the value of $\ell_c$  \label{subsec:beta-c-pc} }

From \eqref{eq:pc-eff-t-no-L0}, one can see that the minimum value
of $p_{c}$ happens at $t=0$ for which $p_{c}$ becomes 
\begin{equation}
p_{c}^{\textrm{min-GUP}}=\gamma GM\ell_{c}.
\end{equation}
It is seen that two free parameters contribute to such a minimum value:
the LQG Barbero-Immirzi parameter $\gamma$ and the GUP minimal length
scale $\ell_{c}$. Hence the existence of such a minimum $p_{c}$
is purely quantum gravitational, due to the dependence on the mentioned
parameters. We can go further and compare this minimum value with
the value derived in \cite{Corichi:2015xia} in the framework of LQG.
There, the minimum value of $p_{c}$ for LQG was found to be 
\begin{equation}
p_{c}^{\textrm{min-LQG}}=\gamma GM\sqrt{\Delta},
\end{equation}
where $\Delta$ is the minimum of the area in LQG which is proportional
to the Planck length squared $\ell_{p}^{2}$. If one identifies $p_{c}^{\textrm{min-GUP}}=p_{c}^{\textrm{min-LQG}}$,
then one would obtain 
\begin{equation}
\ell_{c}^{2}=\Delta.\label{eq:beta-c-Delta}
\end{equation}
Assuming such an identification, one can even go further and derive
a relation between $\beta_{c}$ and $\mu_{c}$, the polymer parameter
associated to the radial direction in LQG. Using \eqref{eq:beta-c-Delta},
our prescription \eqref{eq:prescript} and the prescription presented
in \cite{Corichi:2015xia} for the removal of the dependency on $L_{0}$
which reads 
\begin{equation}
\mu_{c}=\frac{\sqrt{\Delta}}{L_{0}},
\end{equation}
one can deduce
\begin{equation}
-\beta_{c}=\mu_{c}^{2}.
\end{equation}
The above value of $\ell_{c}^{2}\propto\ell_{p}^{2}$ is quite small
as expected and consequently one can conclude that the effective corrections
kick in very close to the singularity and become dominant in that
region, while for the majority of the interior the behavior mimics
the classical solutions. We can assume that $\beta_{b}$ or a related
scale also takes on an approximately same value as $\beta_{c}$, given
that one expects $\beta_{b}$ to be of the same order of magnitude
as $\beta_{c}$.

\subsubsection{Modification to the behavior at the horizon \label{subsec:Modif-Horiz}}

Considering the classical solutions and using the prescription \eqref{eq:prescript},
the classical values of the canonical variables at the horizon are,
\begin{align}
b= & 0,\\
p_{b}= & 0,\\
c= & \mp\frac{\ell_{c}}{\sqrt{-\beta_{c}}}\frac{\gamma}{4GM},\\
p_{c}= & 4G^{2}M^{2}.
\end{align}
The effective solutions, however, take on modified values at the horizon.
To first order of Taylor expansion in $\beta_{i}$, these values are
\begin{align}
b= & \pm\sqrt{\beta_{b}}\gamma^{2}\sqrt{-2\ln(\gamma)},\\
p_{b}= & 2GM\gamma\ell_{c}\sqrt{2\frac{\beta_{b}}{\beta_{c}}\ln(\gamma)},\label{eq:pb-corr}\\
c= & \mp\frac{\ell_{c}}{\sqrt{-\beta_{c}}}\frac{\gamma}{4GM}\left(1-\frac{\gamma^{2}\ell_{c}^{2}}{32G^{2}M^{2}}\right),\\
p_{c}= & 4G^{2}M^{2}-\frac{\gamma^{2}}{8}\ell_{c}^{2}.
\end{align}
From here one can see the effective model we introduced also affects
the dynamics near the horizon, albeit to a very small degree, given
the shear small values of $\ell_{c}$ being proportional to the Planck
length. It is clear from the expressions above that the modifications
to the horizon is not only affected by $\ell_{c}$ but also by the
Barbero-Immirzi parameter $\gamma$. Such modifications to the behavior
at the horizon have also been reported in LQG (see, e.g., \cite{Chiou:2008nm}).
Such a scenario is also suggested in other approaches such as the
firewall proposal \cite{Almheiri:2012rt}. Note that in the above
analysis, for the physical quantity $p_{c}$, the corrections at the
horizon are independent of the mass of the black hole and the fiducial
$L_{0}$, and are proportional to $\ell_{c}^{2}$. The correction
term for $c$ has a dependence on $M$ but it decreases with $M$.
Hence in low curvature regimes this terms becomes smaller as expected.
The correction term for $p_{b}$ however, seems to be proportional
to $M$, which means that it is larger at the horizon of more massive
black holes which seems counter-intuitive. However, one should note
that in the physical quantities we considered here such as the expansion
and shear scalar and certainly the minimum nonzero radius for the infalling 2-spheres, $p_{b}$
plays no role. Furthermore, assuming that $\beta_{b}$ and $\beta_{c}$
being of the same order of magnitude, the correction term for $p_{b}$
would be quite small even for large black holes given the presence
of $\ell_{c}^{2}$. More importantly, if $p_{b}$ only shows up in
physical quantities in a combination such that its effect is suppressed
by a factor of $1/M$ and higher inverse powers of $M$, the proportionality
of the corrections to $p_{b}$ to $M$ in \eqref{eq:pb-corr} would
be of no concern.


\section{Conclusion and outlook\label{sec:Conclusion-and-outlook}}

In this work, we have considered the interior of the black hole written
in terms of Ashtekar-Barbero connection variables. However, instead
of applying the loop quantization techniques, we have considered a
deformation of the canonical algebra. Specifically, such deformation
was inspired by the modification of the commutation relations of the
fundamental operators in GUP models. This way we obtained an effective
Hamiltonian which contains quantum effective corrections to the classical
Hamiltonian. These corrections are accompanied by the introduction
of two GUP parameters $\beta_{b},\,\beta_{c}$ which set the scale
at which such corrections kick in.

Using the effective Hamiltonian, we have derived the equations of
motion for the canonical variables, particularly for $p_{c}$ which
represents the square of the radius of the infalling 2-spheres in
the interior of the black hole. 
We showed that there is a minimum
value for $p_{c}$ and it does not vanish anywhere inside the black
hole. Hence, since all the Riemann invariants for this model are proportional
to $\frac{1}{p_{c}^{n}}$, with $n>0$, there will be no physical
singularity in the interior and the classical singularity is resolved.
Such a singularity resolution has also been reported
in model analyzed using LQG techniques \cite{Bojowald:2004af,Ashtekar:2005qt,Bojowald:2005cb,Bohmer:2007wi,Boehmer:2008fz,Corichi:2015xia,BenAchour:2017ivq,Ashtekar:2018cay,BenAchour:2018khr,Alesci:2018loi,Barrau:2018rts,Alesci:2019pbs,Aruga:2019dwq,BenAchour:2020gon,Bodendorfer:2019cyv,Bodendorfer:2019nvy,Bojowald:2008bt,Bojowald:2008ja,Bojowald:2016itl,Bojowald:2016vlj,Bojowald:2018xxu,Brahma:2014gca,Campiglia:2007pb,Chiou:2008nm,Corichi:2015vsa,Cortez:2017alh,Gambini:2008dy,Gambini:2009ie,Gambini:2011mw,Gambini:2013ooa,Gambini:2020nsf,Husain:2004yz,Husain:2006cx,Kelly:2020lec,Kelly:2020uwj,Kreienbuehl:2010vc,Modesto:2005zm,Modesto:2009ve,Olmedo:2017lvt,Thiemann:1992jj,Zhang:2020qxw,Ziprick:2016ogy,Campiglia:2007pr,Gambini:2009vp,Rastgoo:2013isa,Corichi:2016nkp,Morales-Tecotl:2018ugi,Blanchette:2020kkk}.
However, to our knowledge, our present work is one of the first of
its kind that shows that both the singularity resolution and the emergence of a minimum nonzero radius for the infalling 2-spheres
also happen explicitly in at least some of the models inspired by GUP-type deformed algebra.

Additionally, we also introduced a prescription to cure the dependence
of physical quantities of interest on the fiducial length $L_{0}$.
Such a prescription leads to the introduction of a physical minimum
length scale $\ell_{c}$. By identifying the monimum value of $p_{c}$
we derived here and the one derived in the framework of LQG in \cite{Corichi:2015xia},
we found $\ell_{c}^{2}=\Delta$ where $\Delta$ is the minimum of
the area in LQG. The same line of reasoning also leads to the relation
$-\beta_{c}=\mu_{c}^{2}$ between $\beta_{c}$ and the radial polymer
parameter $\mu_{c}$ in LQG. 

We also briefly studied the effects of such GUP-type modification
on the behavior at the horizon. Our results show that although the
dynamics at the horizon will be different from the classical GR, nevertheless,
due to the small value of $\beta_{b},\,\ell_{c}$, these modifications
will be quite small. 

In a future work, we will study the modifications to the dynamics
of the interior of the Schwarzschild black hole, not by deforming
the classical algebra, but by quantizing the model and then deforming
the quantum commutation relations \cite{preparation}. We will then
study the state and the spectrum of the relevant operators that reveal
the dynamics of the interior and the fate of the region which used
to be the classical singularity.

\acknowledgments{

OO thanks the support of CONACYT Project 257919, UG Project CIIC 188/2019 and PRODEP  (Professional Program for
Teachers Development, Guanajuato). WY would like to thank the financial support granted by CONACYT (National Council of Science and Technology, Mexico).
S. R. acknowledges the support from the Natural Sciences and Engineering Research Council of Canada (NSERC) Discovery Grant RGPIN-2021-03644 and DGECR-2021-00302. 
}

\bibliographystyle{JHEP}
\bibliography{References}

\end{document}